\documentclass[conference]{IEEEtran}
\usepackage{array}
\usepackage[pdftex]{graphicx,xcolor}
\usepackage{xcolor}
\usepackage[T1]{fontenc}
\usepackage{lmodern}
\usepackage{textcomp}
\usepackage{latexsym}
\usepackage[fleqn]{amsmath}
\usepackage{amssymb}
\usepackage{bm}

\begin{document}

\title{Shifted Coded Slotted ALOHA}

\author{
  \IEEEauthorblockN{Tomokazu Emoto and Takayuki Nozaki}
  \IEEEauthorblockA{
    Dept. of Informatics,     Yamaguchi University\\
    1677-1 Yoshida, Yamaguchi-shi, Yamaguchi 753-8512 JAPAN\\
    Email: \tt{\{i005vb,tnozaki\}@yamaguchi-u.ac.jp}}
}

\maketitle

\begin{abstract}
The random access scheme is a fundamental scenario in which users transmit through a shared channel and cannot coordinate each other.
In recent years, successive interference cancellation (SIC) was introduced into the random access scheme.
It is possible to decode transmitted packets using collided packets by the SIC.
The coded slotted ALOHA (CSA) is a random access scheme using the SIC.
The CSA encodes each packet using a local code prior to transmission.
It is known that the CSA achieves excellent throughput.
On the other hand, it is reported that in the coding theory time shift improves the decoding performance for packet-oriented erasure correcting codes.
In this paper, we propose a random access scheme which applies the time shift to the CSA in order to achieve better throughput.
Numerical examples show that our proposed random access scheme achieves better throughput and packet loss rate than the CSA.
\end{abstract}

\IEEEpeerreviewmaketitle

\section{Introduction}
The random access scheme is a fundamental scenario in which users transmit through a shared channel and cannot coordinate each other.
A simple random access scheme is ALOHA \cite{abramson1970ALOHA} proposed by Abramson.
In ALOHA, users transmit their packets whenever they want.
If two or more packets collide, the users re-transmit their packets after waiting random time.
Roberts proposed slotted ALOHA \cite{roberts1975ALOHA}, in which users are synchronized and the transmission starts at the beginning of time slots.

Casini et al.\ \cite{casini2007contention} proposed contention resolution diversity slotted ALOHA (CRDSA).
In CRDSA, each user transmits two copies of the packet.
If two or more packets collide, the receiver re-solves the packets via successive interference cancellation (SIC) \cite{cover1972broadcast}.
Roughly speaking, in the SIC if several packets are transmitted simultaneously, the receiver gets the sum of those packets.
Hence, the random access schemes with the SIC (e.g, CRDSA) can be regarded to as packet-oriented erasure correcting coding systems. 
Liva \cite{liva2011graph} proposed irregular repetition slotted ALOHA (IRSA) as a generalization of CRDSA to improve the throughput.
From the coding theoretic aspect, the factor graph of IRSA is regarded to as the Tanner graph of an irregular low-density parity-check (LDPC) code \cite{Gallager_LDPC,luby2001improved}.
Paolini et al.\ \cite{paolini2015coded} proposed coded slotted ALOHA (CSA) from the construction of a doubly generalized LDPC code \cite{wang2006doubly} as a generalization of the IRSA.

Nowadays, it is known that packet-oriented erasure correcting codes with shift operations have good decoding performance \cite{dai2017new,nozaki2017zigzag}.
The codes with the shift operations are efficiently decoded by zigzag decoding \cite{gollakota2008zigzag}.

In this paper, we propose a random access scheme, referred to as {\it shifted coded slotted ALOHA (SCSA)}.
The SCSA is a protocol combining the CSA with the shift operations.
This paper gives the procedure of the transmitter and the receiver and compares the throughput of the SCSA with the CSA by numerical examples.

This paper is organized as follows.
Section \ref{sec:pre} briefly introduces the CSA and time shift.
Section \ref{sec:SCSA} proposes the SCSA.
In Section \ref{sec:num}, numerical examples show that the proposed protocol outperforms the CSA in terms of the throughput and the packet loss rate.
Section \ref{sec:con} concludes the paper.


\section{Preliminaries\label{sec:pre}}
In this section, we introduce the system model assumed in this paper.
Moreover, this section briefly introduces the procedure of CSA.
Furthermore, we explain the time shift operation.
The notations given in this section are used throughout the paper.

\subsection{System Model}
In this paper, we assume that $N$ users transmit packets to a receiver.
All the users and the receiver are frame and slot synchronous.
All the users are within the range of detectability of the receiver.
Each user attempts at most one packet transmission per frame.

If several packets are transmitted simultaneously, packet collision occurs.
Collisions are always detected by the receiver.
We assume interference cancellation of the SIC is ideal.
In other words, collisions can be regarded as the sum of the packets.

\subsection{Coded Slotted ALOHA}
The CSA is a protocol in which a packet of each user is encoded prior to transmission in a frame.
\subsubsection{Preliminaries of CSA}
We consider a slotted random access protocol where slots are grouped in the frames.
The time duration of the frame is $T_f$.
The frame is composed of $M$ slots of duration $T_s = T_f/M$.
Every slot in the frame is divided into $k$ slices.
The slot is composed of $k$ slices of duration $T_s/k$.

The packet of a user is divided into $k$ data segments, all of the same length.
The transmission of a segment is enforced within one slice.

\subsubsection{Local Component Codes $\mathcal{C}$ and Code Distribution}
The $k$ segments are encoded by the user via a packet-oriented linear block code, generating $n$ encoded segments, all of the same length.
For each transmission, the $(n,k)$ code is randomly chosen by the user from a set $\mathcal{C} = \{C_1, C_2, \dots, C_{\theta}\}$ of $\theta$ component codes.
Note that the set $\mathcal{C} $ is also known to the receiver.
The component code $C_h$ has length $n_h$, dimension $k$, and minimum distance $d_h \leq 2$.
Moreover, it is a proper code, i.e, it has no idle symbols.

Each user randomly and independently chooses a component code from the set $\mathcal{C}$ according to the code distribution $\Lambda (x) = \sum_{i} \Lambda_i x^i$.
More precisely, the component code is chosen as $C_i$ with probability $\Lambda_i$.

\subsubsection{Encoding Procedures}
The system parameters for the CSA are the number of segments $k$, the set $\mathcal{C}$ of the component codes and the code distribution $\Lambda(x)$.
The encoding procedure of the CSA is as follows.
\begin{enumerate}
\item[(1)]The user divides a packet into $k$ segments.
\item[(2)]The user randomly picks a component code $C_h$ from the set $\mathcal{C}$ according to a code distribution $\Lambda(x)$.
\item[(3)]The user encodes $k$ segments via a component code $C_h$, generating $n_h$ encoded segments.
\item[(4)]The user randomly selects $n_h$ slices in the frame.
\item[(5)]The user equips each encoded segment with protocol-control information, namely, the component code picked by the user and the position of the other $n_h-1$ encoded segments in the frame.
\item[(6)]The user encodes the encoded segments via a physical layer code and transmits those at chosen slices.
\end{enumerate}

\subsubsection{Decoding Procedures}
On the receiver side decoding is performed as follows.
\begin{enumerate}
\item[(1)] \label{stp:dec_CSA1} 
  The receiver searches segments in {\it{clean}} slices (i.e, segments not experiencing collisions).
\item[(2)] 
  The receiver extracts protocol-control information, i.e, the relevant user, the code $C_h$ adopted by the user, and the position of the other $n_h-1$ segments in the frame from the received segments.
\item[(3)] 
  The receiver performs maximum a posteriori (MAP) erasure decoding of the component code $C_h$ adopted by the user in order to recover as many encoded segments as possible.
\item[(4)] \label{stp:dec_CSA4}
  The receiver exploits the recovered segments in order to subtract their contribution of interference in those slices where collisions occur.
\item[(5)] This procedure from Step 1 to Step 4 is iterated until either all slices have been cleaned or collisions persist but no further encoded segments can be recovered via MAP erasure decoding.
\end{enumerate}

\subsection{Time Shift}
This section gives the key idea of this paper, namely shift operation or time shift to the segments.
From the sum of shifted segments, the source segments can be recovered by zigzag decoding\cite{dai2017new,nozaki2017zigzag}.
This section explains the time shift to a segment and zigzag decoding via a toy example.

We assume that there are two source segments $s_1 = (s_{1,1}, s_{1,2}, \dots, s_{1,\ell}), s_2 = (s_{2,1}, s_{2,2}, \dots, s_{s,\ell})$ of length $\ell$.
From those source segments, we generates two ``encoded'' segments $c_1, c_2$.
The first encoded segment $c_1 = (c_{1,1}, c_{1,2}, \dots, c_{1,\ell})$ is generated from the addition of two source segments $s_1$ and $s_2$.
The second encoded segment $c_2 = (c_{2,1}, c_{2,2}, \dots, c_{2,\ell}, c_{2,\ell+1})$ is generated from the addition of $s_1$ and $s_2$ with a right shift.
After shifting the packet, zeros are filled.

For this example, the zigzag decoding algorithm proceeds as the following way. 
The decoder has recovered $s_{1,1}$ from $c_{2,1}$ since $s_{1,1} = c_{2,1}$.
The decoder has recovered $s_{2,1}$ by solving $c_{1,1} = s_{1,1} + s_{2,1} = c_{2,1} + s_{2,1}$.
Similarly, the decoder has recovered $s_{1,2}, s_{2,2}, s_{1,3}, s_{2,3}, \dots, s_{1,\ell}, s_{2,\ell}$ and decoding is success.


\section{Shifted Coded Slotted ALOHA\label{sec:SCSA}}
In this section, we propose a protocol which applies the time shift to the CSA.
Recall that the CSA is provided by the three system parameters $(k, \mathcal{C}, \Lambda(x))$.
On the other hand, the system parameters for the SCSA are $k, \mathcal{C}, \Lambda(x), \Delta(x)$, i.e, the shift distribution $\Delta(x)$ is added to the system parameters of the CSA.

\subsection{Shift Distribution $\Delta(x)$}
In this paper, we refer to the unit of digital information as a word, which is composed of $w$ bits.
Moreover, we assume that the shift operation is performed in word-wise.
In other word, $s$ word-wise shift equals to $sw$ bit-wise shift.
We assume that the length of the segment is $l$ words.

We refer to the number of word-wise shifts as the shift amount.
We denote the maximum number of the shift amount, by $s_{\rm{max}}$.
Let $\mathcal{S}$ be a set of the shift amount, i.e, $\mathcal{S} = \{0, 1, 2, \dots, s_{\rm{max}}\}$.
Each user randomly and independently chooses a shift amount from the set $\mathcal{S}$ according to the shift distribution $\Delta(x) = \sum_{i} \Delta_i x^i$.
More precisely, the shift amount is chosen as $d$ with probability $\Delta_d$.

\subsection{Encoding Procedure of the SCSA}
The encoding procedure of the SCSA is as follows.
\begin{enumerate}
\item[(1)]The user divides a packet into $k$ segments.
\item[(2)]The user randomly picks a code $C_h$ from the set $\mathcal{C}$ according to a code distribution $\Lambda(x)$.
\item[(3)]The user encodes $k$ segments via a $C_h$, generating $n_h$ encoded segments.
\item[(4)]The user randomly selects $n_h$ slices in the frame.
\item[(5)]The user randomly and independently picks $n_h$ shift amounts from the set $\mathcal{S}$ for encoded segments according to a shift distribution $\Lambda(x)$.
\item[(6)]The user equips both beginning and end of each encoded segment with protocol-control information (namely, the component code picked by the user, the position and the shift amount of the other $n_h-1$ encoded segments).
\item[(7)] For a segment whose shift amount is $d$, the user fills $dw$ zeros at the beginning of the segment and $(s_{\rm{max}} - d)w$ zeros at the end of it.
\item[(8)]The user encodes the encoded segments via a physical layer code and transmits those at chosen slices.
\end{enumerate}

\subsection{Decoding Procedure of the SCSA}
On the receiver side decoding of the SCSA is performed as follows.
\begin{enumerate}
\item[(1)] \label{stp:dec_SCSA1}
  The receiver searches segments in clean slices.
\item[(2)]
  The receiver extracts protocol-control information (i.e, the code $C_h$ adopted by the user, and the position and the shift amount of the other $n_h-1$ encoded segments) from the received segments.
\item[(3)]
  The receiver performs MAP erasure decoding of $C_h$ in order to recover as many encoded segments as possible.
\item[(4)] \label{stp:dec_SCSA4}
  The receiver exploits the recovered segments in order to subtract their contribution of interference in those slices where collisions occur.
\item[(5)]
  This procedure from Step 1 to Step 4 is iterated until either all slices have been cleaned or collisions persist, but no further encoded segments can be recovered via MAP erasure decoding.
\item[(6)] \label{stp:dec_SCSA6}
  The receiver searches segments in partly clean slices (i.e, words of segments not experiencing collisions).
\item[(7)]
  The receiver extracts protocol-control information from the beginning or the end word of the received segments.
\item[(8)]
  The receiver performs MAP erasure decoding of $C_h$ in order to recover as many words of the encoded segment as possible.
\item[(9)] \label{stp:dec_SCSA9}
  The receiver exploits the recovered word of the encoded segment in order to subtract their contribution of interference in those slices where collisions occur.
\item[(10)]
  This procedure from Step 6 to Step 9 is iterated until either all slices have been cleaned or collisions persist but no further the words of encoded segments can be recovered via MAP erasure decoding.
\end{enumerate}

\subsection{Example of the SCSA}
\begin{figure}[tb]
  \centering
  \includegraphics[width=7cm]{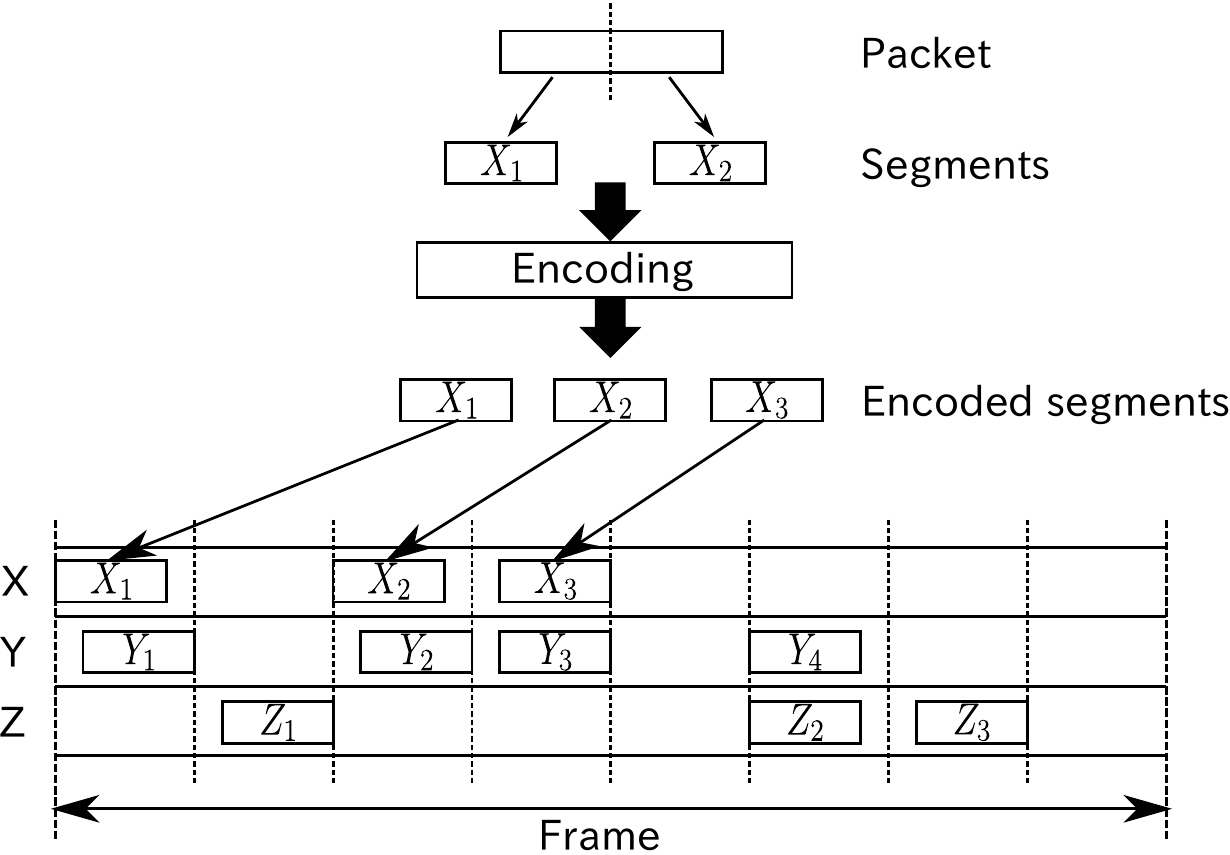}
  \caption{An example of the SCSA \label{fig:propose}}
\end{figure} 

In Fig.~\ref{fig:propose} a pictorial representation of the encoding and transmission process is provided in the case of $N = 3$ users (indexed as user X, user Y, user Z), $kM = 8$ slices (indexed from 1 to 8), and $s_{\rm{max}} = 1$. 
A set $\mathcal{C}$ is $\{C_1,C_2\}$ and a generator matrix $\mathbf{G}_i$ of code $C_i$ is
\begin{equation}
  \mathbf{G}_1 =
  \begin{bmatrix}
    101 \\ 011
  \end{bmatrix}, 
  \quad
  \mathbf{G}_2 =
  \begin{bmatrix}
    1110 \\ 0111
  \end{bmatrix}.
\end{equation}

\subsubsection{Encoding}
Each packet is split into $k = 2$ information segments.
Out of the three users, the users X and Z employ a code $C_1$ while the user Y employs a code $C_2$. 
The user X performs systematic encoding of its two data segments, generating one parity segment.
The three segments are then transmitted into the frame slices of indexes 1, 3, 4 with the shift amount 0, 0, 1.
The encoded segments of the users Y and Z (performing systematic encoding as well) are transmitted in slices of indexes 1, 3, 4, 6 and 2, 6, 7 with the shift amount 1, 1, 1, 0 and 1, 0, 1, respectively.

\subsubsection{Decoding}
A collision is detected by the receiver on the slices with indexes 1, 3, 4, and 6, while interference-free segments are received on the slices with indexes 2 and 7.
It is easy to recognize that MAP erasure decoding of the code $C_1$ employed by the user Z allows us to recover the one missing segment of this user.
The contribution of interference of this one segment can then be subtracted from the corresponding slice (of index 6), cleaning the segment transmitted by the user Y.
The remaining segments cannot be recovered by the decoding procedure of the CSA.
Hence, the receiver starts zigzag decoding.

Interference-free words are received in indexes 1 and 3.
MAP erasure decoding of the code $C_1$ employed by the user X allows us to recover the word at the beginning of the segment of this user.
The contribution of interference of this one word can then be subtracted from the corresponding word (of index 3), cleaning the word transmitted by the user Y.
MAP erasure decoding of the code $C_2$ employed by the user Y allows us to recover the words at the beginning of the segment of this user.
The contribution of interference of these two words can then be subtracted from the corresponding word (of indexes 1 and 3), cleaning the words transmitted by the user X.
Similarly, the receiver can recover all the words of the users X and Y, and decoding successes.


\section{Numerical Examples \label{sec:num}}
We perform numerical simulations for frame size $M$ and user population size $N$.

\subsection{System Parameters \label{sec:sp}}
Recall that the SCSA is provided by the four system parameters $(k, \mathcal{C}, \Lambda(x), \Delta(x))$.
We employ the set $\mathcal{C}$ and the code distribution $\Lambda(x)$ which achieve the largest peak throughput in \cite{paolini2015coded} when $k=1, 2$.
We use the set $\mathcal{C}_1$ (resp.~$\mathcal{C}_2$) and the code distribution $\Lambda_1(x)$ (resp.~$\Lambda_2(x)$) when $k=1$ (resp.~$k=2$) detailed in the following.
\begin{align*}
  \mathcal{C}_{1} &= \{C_{1,1},C_{1,2},C_{1,3}\}, \\
  \mathbf{G}_{1,1} &= \begin{bmatrix}11 \end{bmatrix}, \quad
  \mathbf{G}_{1,2}  = \begin{bmatrix}111 \end{bmatrix}, \quad
  \mathbf{G}_{1,3}  = \begin{bmatrix}111111 \end{bmatrix},\\
  \Lambda_1 (x) &= 0.554016x +  0.261312x^2 + 0.184672x^3, \\
  \mathcal{C}_{2} &= \{C_{2,1},C_{2,2},C_{2,3},C_{2,4},C_{2,5},C_{2,6},C_{2,7}\}, \\
  \mathbf{G}_{2,1} &= \begin{bmatrix} 110 \\011 \end{bmatrix}, \quad
  \mathbf{G}_{2,2}  = \begin{bmatrix} 1100 \\1111 \end{bmatrix},\quad 
  \mathbf{G}_{2,3}  = \begin{bmatrix} 11100 \\00111 \end{bmatrix},\\
  \mathbf{G}_{2,4} &= \begin{bmatrix} 11110 \\00011 \end{bmatrix}, \quad
  \mathbf{G}_{2,5}  = \begin{bmatrix} 11111 \\00011 \end{bmatrix}, \\
  \mathbf{G}_{2,6} &= \begin{bmatrix} 11110000000 \\ 00111111111 \end{bmatrix}, \quad
  \mathbf{G}_{2,7} = \begin{bmatrix} 111111110000 \\ 000001111111 \end{bmatrix},\\
  \Lambda_2 (x) &= 
  0.259929x + 0.053247x^2 + 0.259293x^3  \\ &~~~~+ 0.098353x^4 + 0.080412x^5 \\
  &~~~~ + 0.105258x^6 + 0.134509x^7,
\end{align*}
where $\mathbf{G}_{i,j}$ is a generator matrix of $C_{i,j}$.

We employ the uniform distribution for the shift distribution $\Delta(x)$.
More precisely,
\begin{align*}
  {\Delta}(x) = \frac{1}{s_{\rm{max}}+1} \sum^{s_{\rm{max}}}_{d=0}x^d.
\end{align*}

\subsection{Evaluation Criteria}
The normalized offered traffic (or channel traffic) $G$ is given as follows,
\begin{align*}
  G 
  =
  \frac{N}{M}
  \times
  \frac{\ell}{\ell + s_{\rm{max}} + 1 + \mathbb{ I } [s_{\rm{max}} \geq 1]},
\end{align*}
where $\mathbb{I}[s_{\rm{max}} \geq 1]$ is the indicator function which is 1 if the condition in the square brackets is fulfilled and otherwise 0.
Here, the second factor represents the effect of protocol control information and time shift.
The packet loss rate (PLR) and the segment loss rate (SLR) are defined as :
\begin{align*}
  {\rm{PLR}} &= \frac{\#~\rm{unrecovered~packet}}{N}, \\
  {\rm{SLR}} &= \frac{\#~\rm{unrecovered~segment}}{kN},
\end{align*}
where $\#$ stands ``the number of''.
Note that the number of users equals to $N$.

The normalized throughput (or channel output) $T_p$ (resp.~$T_s$) is defined as the probability of successful packet (resp.~segment) transmission per slot (resp.~slice).
More precisely,
\begin{align*}
  T_p =  G(1-{\rm PLR}), \quad 
  T_s = G(1-{\rm SLR})  .
\end{align*}

\subsection{Results}
The simulations carry out at two cases.
The first case of simulations deals with $k = 1$.
We adopt $\mathcal{C}_1, \Lambda(x)_1,$ and $\Delta(x)$ introduced in Section \ref{sec:sp}.
The number of segments equals the number of packets when $k = 1$.
Hence, PLR = SLR and $T_p=T_s$.

The second case of simulations deals with $k = 2$.
We adopt $\mathcal{C}_2, \Lambda(x)_2,$ and $\Delta(x)$ introduced in Section \ref{sec:sp}.
The set of simulations assumes a fixed frame length of $M=500$ and a fixed segment length of $\ell = 200$.

\subsubsection{Case of $k=1$}
Fig.~\ref{fig:k1p} (resp.~Fig.~\ref{fig:k1tp}) compares the PLR (resp.~throughput) of CSA with SCSA.
The simulations are performed with $\mathcal{C}_1$ and $\Lambda_1(x)$.
The curves with $s_{\rm{max}} = 1,2$ give of the SCSA with the maximum shift amount $s_{\rm{max}} = 1,2$, respectively. 
As shown in Fig.~\ref{fig:k1p}, the PLR is monotonically decreasing as $s_{\rm{max}}$ increases.
From Fig.~\ref{fig:k1tp}, the peak throughput is increasing as $s_{\rm{max}}$ increases.
Hence, we conclude that the SCSA outperforms the CSA for $k=1$.

\subsubsection{Case of $k=2$}
\begin{figure}[tb]
  \centering
  \includegraphics[width=.8\linewidth]{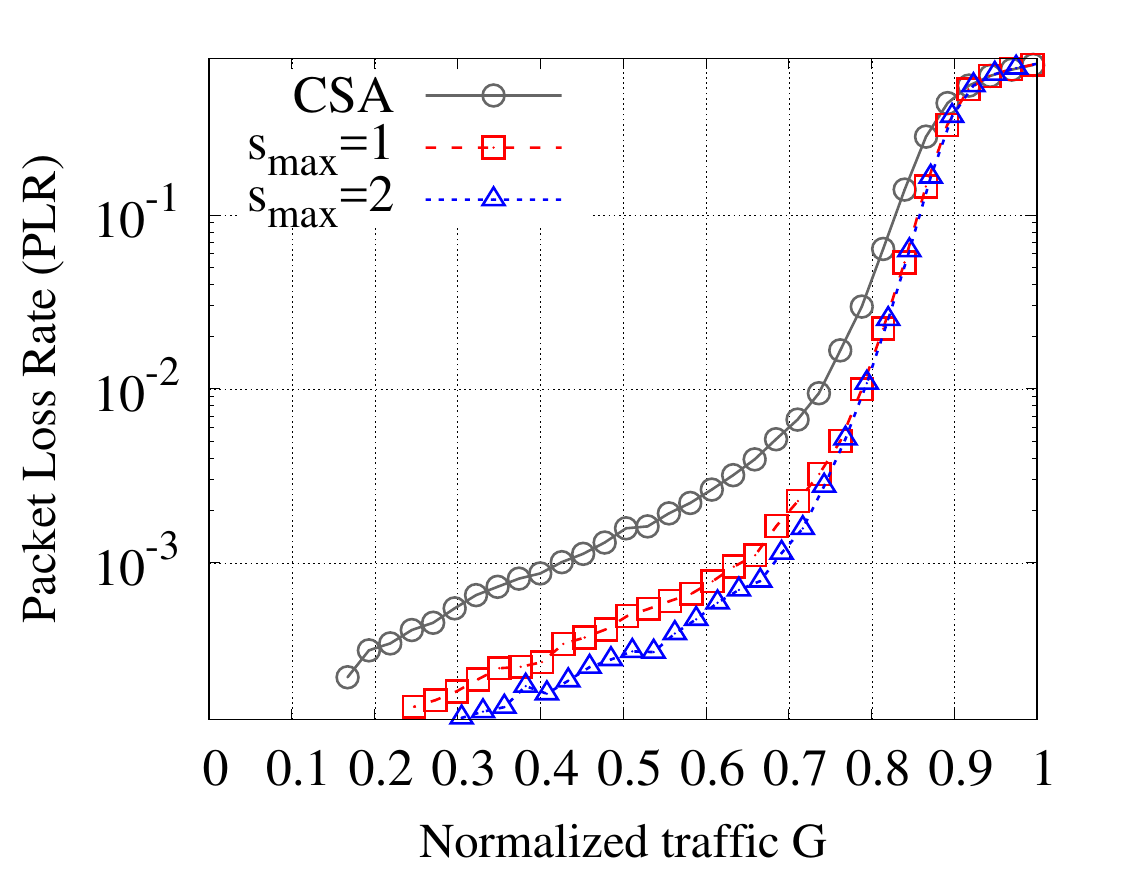}
  \caption{Comparison of packet loss rate of SCSA ($s_{\max} = 1,2$) with CSA for $k=1, M=500, \ell = 200$ \label{fig:k1p}}

  \includegraphics[width=.8\linewidth]{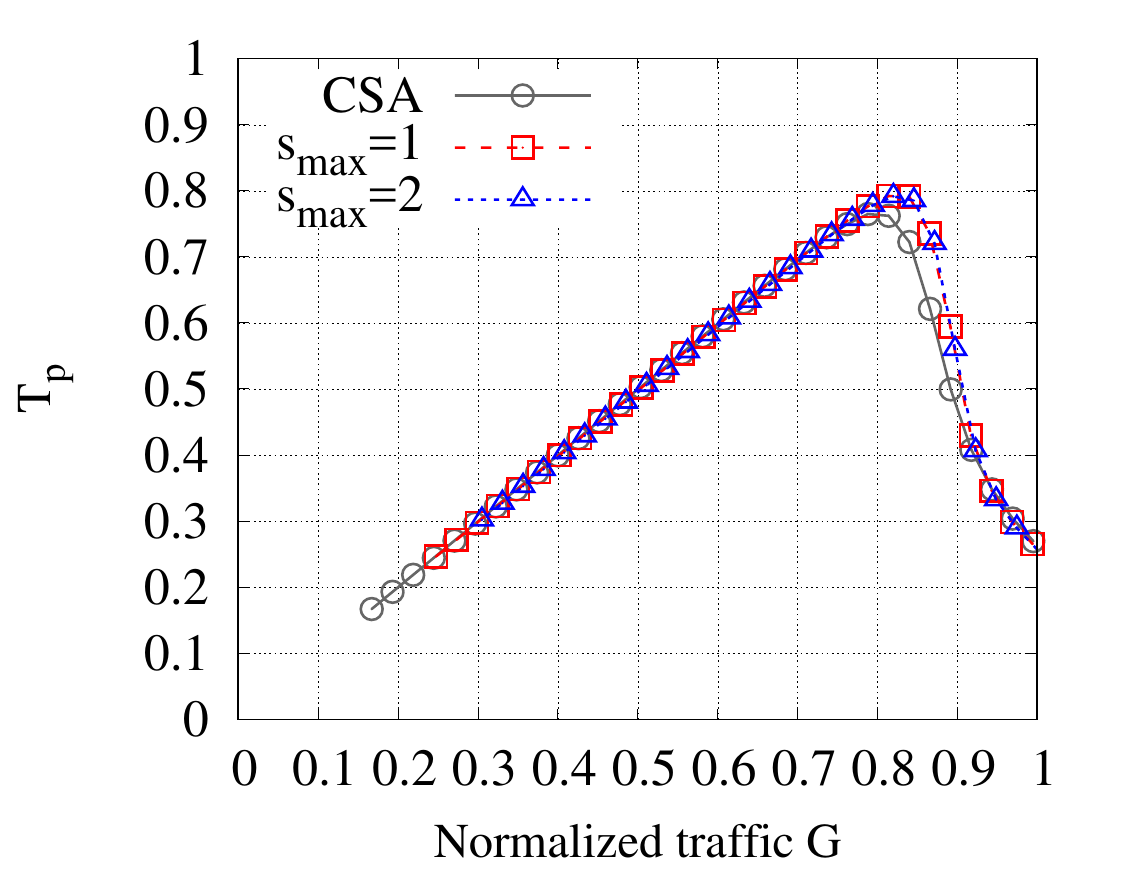}
  \caption{Comparison of normalized throughput of SCSA ($s_{\max} = 1,2$) with CSA for $k=1, M=500, \ell = 200$ \label{fig:k1tp}}

  \includegraphics[width=.8\linewidth]{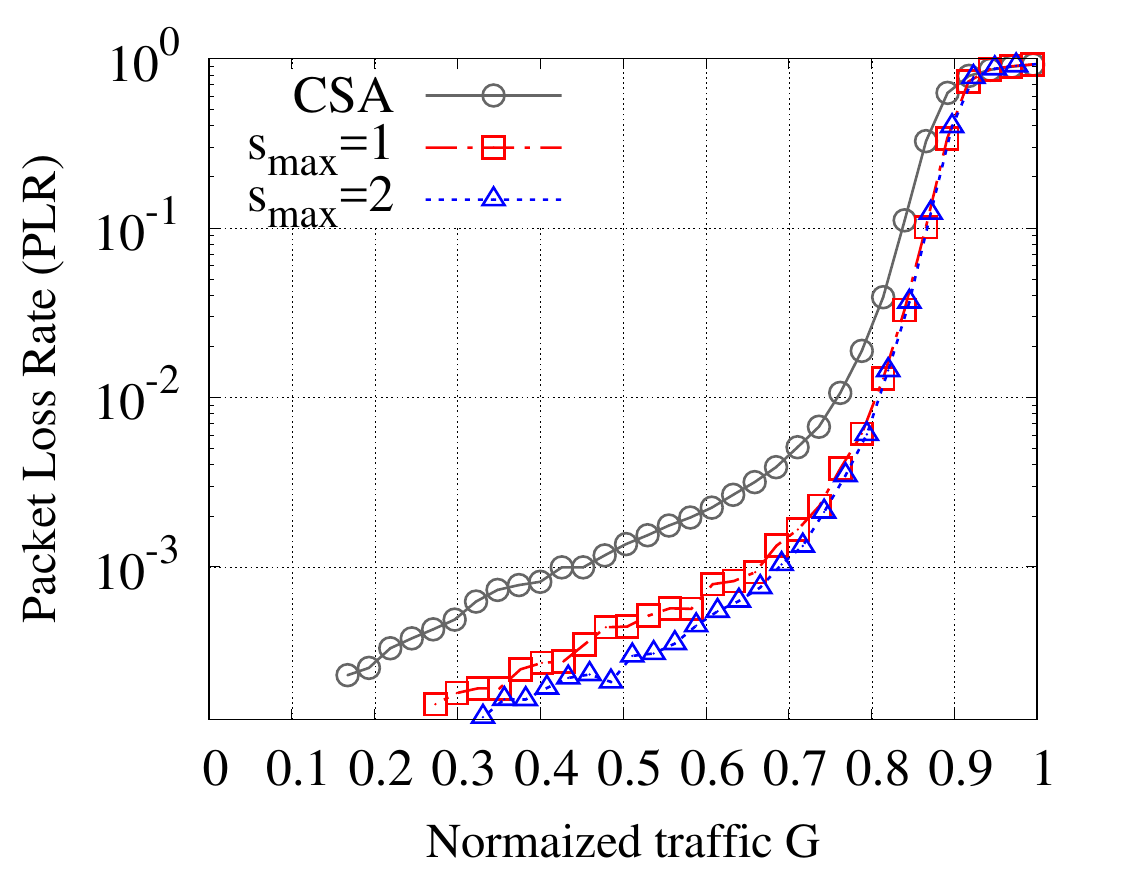}
  \caption{Comparison of packet loss rate of SCSA ($s_{\max} = 1,2$) with CSA for $k=2, M=500, \ell = 200$ \label{fig:k2p}}
\end{figure}

\begin{figure}[tb]
  \centering
  \includegraphics[width=.8\linewidth]{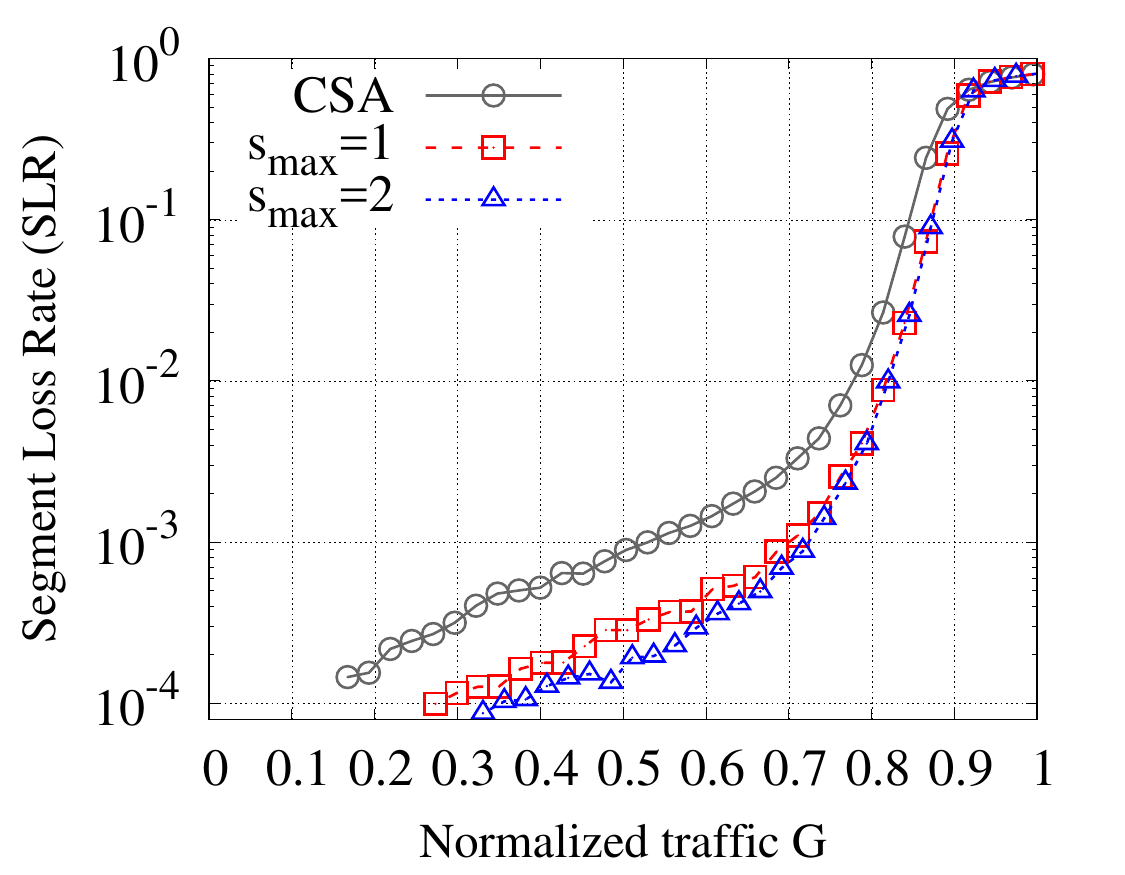}
  \caption{Comparison of segment loss rate of SCSA ($s_{\max} = 1,2$) with CSA for $k=2, M=500, \ell = 200$  \label{fig:k2s}}

  \includegraphics[width=.8\linewidth]{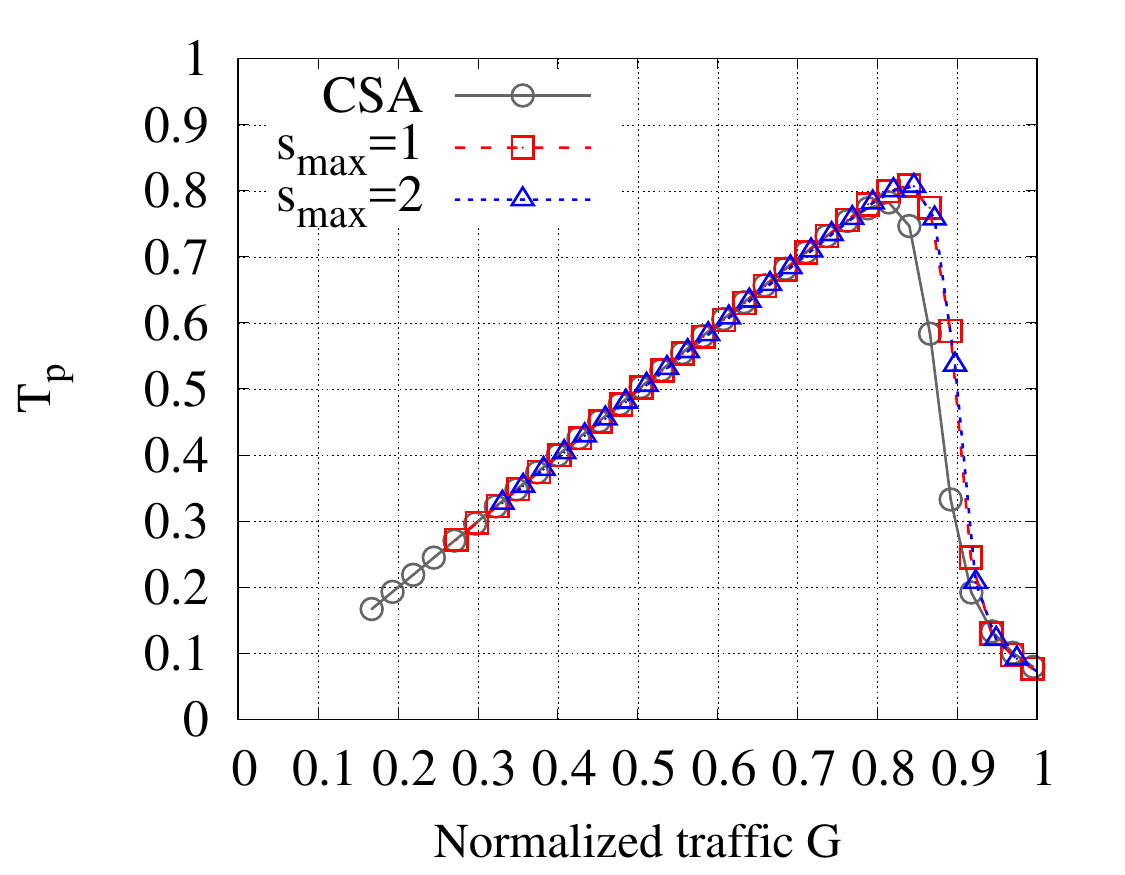}
  \caption{Comparison of normalized throughput $T_{\rm p}$ of SCSA ($s_{\max} = 1,2$) with CSA for $k=2, M=500, \ell = 200$  \label{fig:k2tp}}

  \includegraphics[width=.8\linewidth]{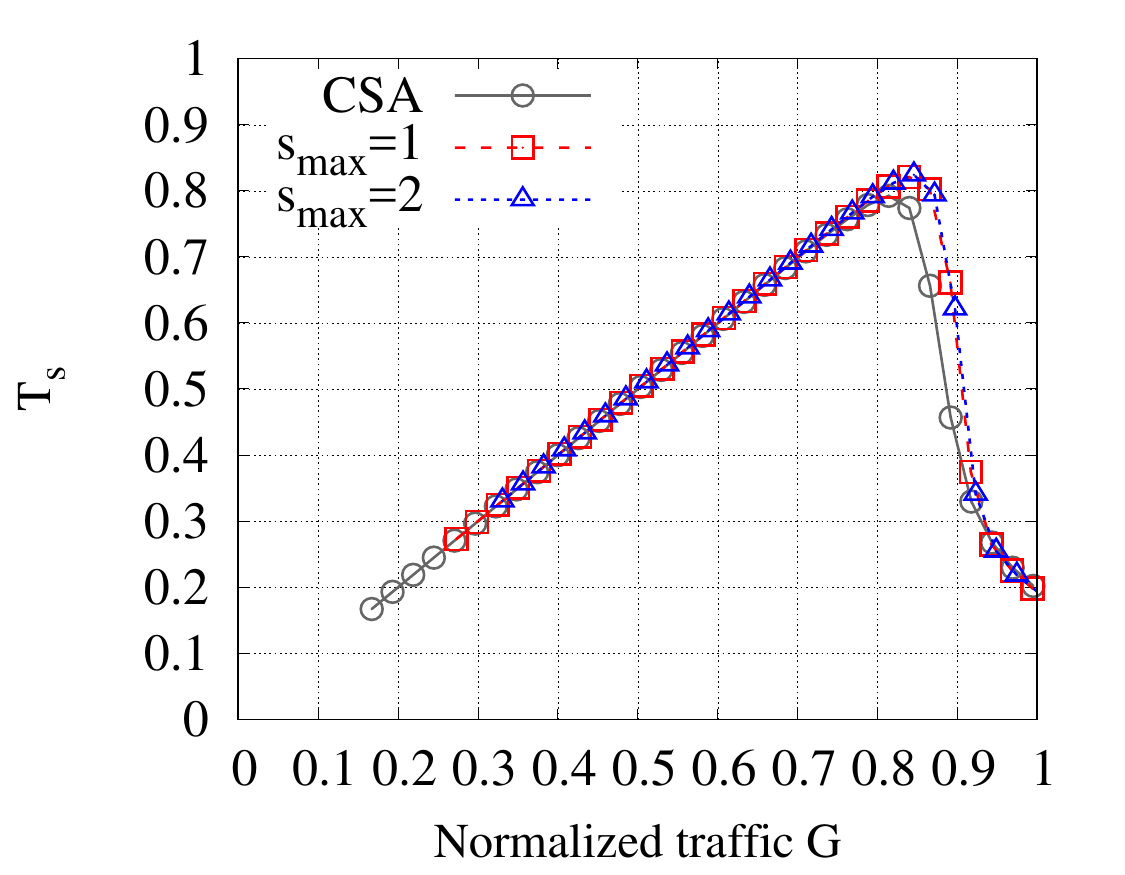}
  \caption{Comparison of normalized throughput $T_{\rm s}$ of SCSA ($s_{\max} = 1,2$) with CSA for $k=2, M=500, \ell = 200$  \label{fig:k2ts}}
\end{figure}

In Figs.~\ref{fig:k2p}, \ref{fig:k2s}, \ref{fig:k2tp} and \ref{fig:k2ts} compare the PLR, the SLR, and the throughput of packet and segment for SCSA with CSA, respectively.
The simulations are performed with $\mathcal{C}_2$ and $\Lambda_2(x)$.
The curves with $s_{\rm{max}} = 1,2$ give of the SCSA with the maximum shift amount $s_{\rm{max}} = 1, 2$, respectively. 
As shown in Figs.~\ref{fig:k2p} and \ref{fig:k2s}, the PLR and the SLR are monotonically decreasing as $s_{\rm{max}}$ increases.
From Figs.~\ref{fig:k2tp} and \ref{fig:k2ts}, the peak throughput of packet and segment are increasing as $s_{\rm{max}}$ increases.
Hence, the SCSA outperforms the CSA for $k=2$.

\section{Conclusion\label{sec:con}}
In this paper, we have proposed the SCSA that is a protocol combining the CSA with the shift operations.
Numerical examples have shown that our proposed protocol achieves better throughput than the CSA.

\end{document}